\documentclass[aip,jcp,reprint,floatfix,showpacs]{revtex4-1}

\usepackage{amssymb}
\usepackage{amsmath}
\usepackage{amsfonts}
\usepackage{graphicx}

\newcommand{\fancy}{\mathcal}

\newcommand{\FE}{\kappa}

\newcommand{\FW}{\fancy{W}}

\renewcommand{\vec}[1]{\boldsymbol{#1}}

\newcommand{\mat}[1]{\mathbb{#1}}

\newcommand{\beq}{\begin{eqnarray}}
\newcommand{\eeq}{\end{eqnarray}}

\newcommand{\tr}{\text{Tr}}

\newcommand{\half}{\frac{1}{2}}

\newcommand{\rcite}[1]{Ref.~\onlinecite{#1}}

\newcommand{\rcites}[1]{Refs~\onlinecite{#1}}

\newcommand{\Hx}{\text{Hx}}

\newcommand{\xrm}{\text{x}}
\newcommand{\crm}{\text{c}}

\newcommand{\xc}{\text{xc}}
\newcommand{\Hxc}{\text{Hxc}}

\newcommand{\EHx}{{\cal E}_{\Hx}}
\newcommand{\Ec}{{\cal E}_{\crm}}

\newcommand{\Ts}{{\cal T}_{s}}

\newcommand{\E}{{\cal E}}
\newcommand{\F}{{\cal F}}

\newcommand{\Psinl}{\Psi^{n,\lambda}}

\ifdefined\vr
\renewcommand{\vr}{\vec{r}}
\else
\newcommand{\vr}{\vec{r}}
\fi

\newcommand{\iket}[1]{|#1\rangle}
\newcommand{\ibraket}[2]{\langle#1|#2\rangle}
\newcommand{\ibraketop}[3]{\langle#1|#2|#3\rangle}
\newcommand{\ibkouter}[1]{|#1\rangle\langle#1|}

\newcommand{\iout}{\ibkouter}

\newcommand{\EEXX}{{\text{EEXX}}}

\newcommand{\gs}{\text{gs}}
\newcommand{\ts}{\text{ts}}
\renewcommand{\ss}{\text{ss}}
\newcommand{\st}{\text{ss--ts}}

\newcommand{\np}{n^{(p)}}

\newcommand{\up}{\mathord{\uparrow}}
\newcommand{\down}{\mathord{\downarrow}}

\newcommand{\nh}{\hat{n}}
\newcommand{\Th}{\hat{T}}
\renewcommand{\th}{\hat{t}}
\newcommand{\Wh}{\hat{W}}
\newcommand{\vh}{\hat{v}}

\newcommand{\Hh}{\hat{H}}

\newcommand{\Gammah}{\hat{\Gamma}}

\usepackage[normalem]{ulem}
\usepackage{color}
\usepackage{xcolor}

\definecolor{Mygrey}{gray}{0.80}

\begin{document}
\title{Charge transfer excitations from exact and approximate ensemble Kohn-Sham theory}
\author{Tim Gould}\affiliation{Qld Micro- and Nanotechnology Centre, %
  Griffith University, Nathan, Qld 4111, Australia}
\author{Leeor Kronik}\affiliation{Department of Materials and
  Interfaces, Weizmann Institute of Science, Rehovoth 76100, Israel}
\author{Stefano Pittalis}\affiliation{CNR-Istituto di Nanoscienze, Via
  Campi 213A, I-41125 Modena, Italy}
\begin{abstract} 
  By studying the lowest excitations of an exactly solvable
  one-dimensional molecular model, we show that components of
  Kohn-Sham ensembles can be used to describe charge transfers.
  Furthermore, we compute the approximate excitation energies obtained
  by using thee exact ensemble densities in the recently
  formulated ensemble Hartree-exchange theory
  [Gould and Pittalis, {\em Phys. Rev. Lett.} 119, 243001 (2017)].
  Remarkably, our results show that triplet excitations
  are accurately reproduced across a dissociation curve in
  all cases tested, even in systems where ground state energies are
  poor due to strong static correlations.
  Singlet excitations exhibit larger deviations from exact results but
  are still reproduced semi-quantitatively.
\end{abstract}
\maketitle

\section{Introduction}

Density functional theory\cite{HohenbergKohn,KohnSham} (DFT)
is a widely employed approach to the many-electron problem, which has
proven to be immensely useful for studying a wide range of issues in
chemistry and physics. DFT is inherently a ground state theory, but its
time-dependent counterpart (TDDFT)\cite{Runge84} is an
increasingly important tool for the study of excited-state properties.

\begin{figure}
  \includegraphics[width=\linewidth]{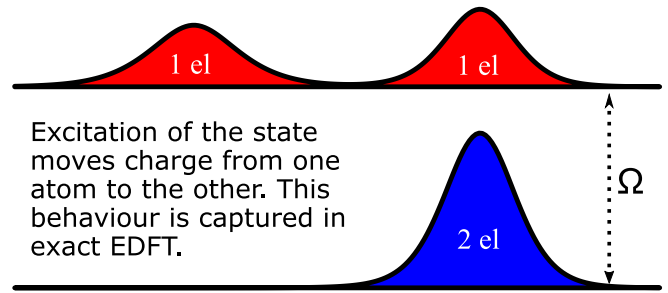}
  \caption{An illustration of charge transfer in a dimer, from a ground state
    with two electrons on the right atom to an excited state with
    one electron on each atom. $\Omega$ is the difference in energy of
    the two states considered, i.e., the excitation energy.
   \label{fig:Fig0}}
\end{figure}

Charge transfer (CT) excitations (illustrated in
Figure~\ref{fig:Fig0}) are physically important phenomena that are
involved in key processes for energy, e.g., photosynthesis,
photovoltaic energy conversion, and
photocatalysis.\cite{DeCastro-CT,Kumar-CT,Hedley-CT}
However, they pose a significant challenge for conventional DFT and
TDDFT approximations.\cite{Maitra2017-tddftReview,Kummel2017-CT} The
fundamental reason behind this challenge is that CT excitations
involve, by definition, transitions between filled states and empty
states with very little spatial overlap. As a consequence, matrix
elements of the exchange-correlation kernel used in linear-response
TDDFT based on Kohn-Sham theory will be vanishly small, and
excitations energies will reduce to Kohn-Sham orbital-energy
difference, unless the kernel exhibits singularity. While the exact
exchange-correlation kernel does indeed exhibit such
behavior,\cite{Thiele2014} standard approximate kernels do not and
typically yield a drastic underestimate of the excitation energy, by
as much as several eV.\cite{Tozer1999}

One useful path to overcome this problem is to capture CT transitions
using constrained DFT.\cite{Kaduk2012-Constrained} However, this
relies on prior knowledge of properties of the chemical system, which
limits its range of applicability and predictive power. 
Optimal tuning\cite{Kronik2012} within generalized Kohn-Sham
theory\cite{Seidl1996} has proven to be highly useful for prediction
of both full and partial CT
excitations.\cite{Kummel2017-CT,Stein2009a,Stein2009b} Still, issues
may arise with strongly heterogeneous systems \cite{Karolewski2013}
and the approach relies inherently on Fock or Fock-like operators,
which can be computationally expensive.
TDDFT calculations within Kohn-Sham theory, based on the
exact-exchange kernel, \cite{Goerling1998-tdEXX,Hellgren2008,Hesselmann2010,
Hellgren2010,Hellgren2012,Bleiziffer2012,Hellgren2013}
can, in principle, capture CT excitations, owing to a highly divergent
kernel. However, this too is computationally intensive and it also
lacks compatible correlation expressions.
Therefore, despite much progress there is still ongoing interest in
developing additional DFT-based strategies that can capture CT
excitations correctly and inexpensively.

One different, low-cost route to the CT problem is afforded by the
Gross, Oliveira, and Kohn
(GOK)\cite{Gross1988-1,Gross1988-2,Oliveira1988-3}
ensemble density functional theory (EDFT),%
\cite{Theophilou1979,Valone1980,Perdew1982,Lieb1983,Savin1996,%
Ayers2006-Axiomatic,Gould2017-Limits}
which offers a statistical ensembles of quantum states that can be
treated similarly to a ground state. EDFT can yield energy differences
directly, as discussed in detail below. Indeed, excited state EDFT has seen
increasing interest of late\cite{Franck2014,Filatov2015,Filatov2016,Deur2017,%
Yang2014,Pribram-Jones2014,Yang2017-EDFT,Gould2017-Limits} as a
potential alternative to TDDFT for excitation
energies. This recent resurgence of GOK EDFT mirrors a growing
interest in more general forms of EDFT, which can deal, e.g., with degenerate
ground states\cite{Lieb1983,Yang2000,Nagy2005,Pittalis2006,Gould2013-LEXX}
and ``open'' systems with a non-integer number of
electrons.\cite{Gould2013-LEXX,Kraisler2013,Kraisler2014,Gorling2015}
Furthermore, a unified EDFT could eventually offer a path to
approximations that can more accurately deal with partitions or
fragments of
systems\cite{Elliott2010,Tang2012,Gordon2012-Frag,Fabiano2014,Nafziger2015}
as bonded fragments will naturally exchange both charge and energy
with their neighbors (i.e. are ``open''), phenomena which require an
ensemble treatment. In light of these potential advantages, it is
important to understand whether exact EDFT has orbitals
and densities that can acquire a direct physical meaning and
are thus amenable to direct approximations, and whether
approximations to EDFT, specifically exact-exchange approximations,
can capture CT excitations. One-dimensional
molecular models provide a convenient test bed to study
the first question. The recently-derived ensemble
Hartree-exchange (Hx) functional, $\EHx[n]$,\cite{Gould2017-Limits}
offers theoretical tools to answer the second
question, as it yields  desirable multi-reference spin-states and
(maximally) ghost interaction free\cite{Gidopoulos2002} energies as an
emergent property of GOK EDFT. 

In this article, we will show that the answer to both questions is a
qualified yes, at least for the cases considered here.
This article is arranged as follows. First, we introduce GOK-EDFT and
its Hartree-exchange approach. Fundamental differences with standard
DFT are spelled-out, too. Next, we describe the model system, present
the results of key tests for the lowest-energy triplet and singlet
excitations, and discuss their significance.
Finally, we summarize and conclude.

\section{Theory}

Conventional DFT uses the electron density $n(\vr)$, rather than the
many-electron wavefunction, as a basic variable. It thereby
makes calculations much more efficient, albeit at the expense of
uncontrolled approximations to the underlying physics. Most DFT
calculations employ the Kohn-Sham formalism\cite{KohnSham},
which involves one-electron orbitals subject to a common potential.
We start our considerations by providing a succinct overview of
standard and ensemble DFT, based on the constrained minimization
approach, introduced and discussed in various forms in
\rcites{Valone1980,Perdew1982,Lieb1983}.

\subsection{Pure-state density functional theory}

Consider a Hamiltonian $\Hh_v=\Th+\Wh+\vh$, where $\Th$ is the
kinetic energy operator, $\Wh$ is the electron-electron interaction
operator and $\vh=\int d\vr v(\vr)\nh(\vr)$ is the interaction
operator for electrons in an external potential $v(\vr)$. The ground
state energy of the Hamiltonian can be found by calculating
$E_0[v]=\min_{\Psi}\ibraketop{\Psi}{\Hh_v}{\Psi}$ subject to
$\ibraket{\Psi}{\Psi}=1$, with $\iket{\Psi}$ a Fermionic
(antisymmetric) wavefunction, i.e., we minimize over wavefunctions.

If we instead use the Levy constrained minimization
approach,\cite{Levy1979} we can transform the process to one where we
find the ground state energy $E_0[v]$ via a minimization over the
one-particle density $n(\vr)$, rather than wavefunctions. This
involves rewriting the minimization as follows:
\begin{align}
  E_0[v]=&\min_{\Psi}\ibraketop{\Psi}{\Hh_v}{\Psi}
  \nonumber\\
  =&\min_{\Psi}\bigg\{ \ibraketop{\Psi}{\Th+\Wh}{\Psi}
  + \int \ibraketop{\Psi}{\nh(\vr)}{\Psi}v(\vr)d\vr \bigg\}
  \nonumber\\
  =&\min_n\bigg\{ \min_{\Psi\to n}\ibraketop{\Psi}{\Th+\Wh}{\Psi}
  + \int n(\vr)v(\vr)d\vr \bigg\}
  \nonumber\\
  \equiv& \min_n\bigg\{ F[n] + \int n(\vr)v(\vr) \bigg\},
  \label{eqn:Evpure}
\end{align}
Here, the intermediate steps define a functional of the particle
density $F[n]$ that
depends only on the form of the kinetic and interaction energy
operators and does not depend on the external potential. The
constraint $\Psi\to n$ in the penultimate expression means the
minimization is taken only over normalized Fermionic wavefunctions
obeying $\ibraketop{\Psi}{\nh}{\Psi}=n(\vr)$, i.e. constrained to the
desired ($N$-representable) density $n(\vr)$.

The  ground state density can be found also by solving 
the ground-state of the Kohn-Sham system.
Kohn-Sham DFT can be viewed from the perspective of the adiabatic
connection,\cite{Harris1984} in which electron-electron interactions
are scaled by $\lambda$. This generalizes the universal
density functional $F[n]$ to
\begin{align}
  F^{\lambda}[n]=&\min_{\Psi\to n}
  \ibraketop{\Psi}{\Th+\lambda\Wh}{\Psi},
  \label{eqn:Fpure}
\end{align}
(again with $\iket{\Psi}$ Fermionic and normalized). 
The constrained minimization in \eqref{eqn:Fpure} can be solved,
for ``typical'' $v$-representable densities $n(\vr)$, by finding the
representative potential $v^{\lambda}[n](\vr)$ for which the ground
state $\iket{\Psinl}$ of $\Hh^{\lambda}=\Th+\lambda\Wh
+\int v^{\lambda}[n] \nh d\vr$ obeys
$n=\ibraketop{\Psinl}{\nh}{\Psinl}$\footnote{This is the definition
of a $v$-representable pure state density $n$.}. In such cases
$v^{\lambda}$ serves as a Lagrange multiplier in the calculation of
$F^{\lambda}$, and thus
$F^{\lambda}[n]=\ibraketop{\Psinl}{\Th+\lambda\Wh}{\Psinl}$.
At full-interaction strength $\lambda=1$, the corresponding potential
$v^1=v$ is simply the external potential of the many-electron
system. With no interactions, $v_s\equiv v^0$ is known as the
Kohn-Sham (KS) potential and, due to the absence of two-body
interactions, and with the exception of degenerate groundstates,
$\iket{\Psi^{n,0}}\equiv \iket{\Phi_s}$ is unambiguously a
{\emph{single} Slater-determinant wavefunction.

From these basic definitions, we can further define two other key
functionals, the non-interacting kinetic energy and the
Hartree-exchange (Hx) functionals:
\begin{align}
  T_s[n]\equiv& F^0[n]=\ibraketop{\Phi_s}{\Th}{\Phi_s}
  \label{eqn:Tspure}
  \\
  E_{\Hx}[n]\equiv&\ibraketop{\Phi_s}{\Wh}{\Phi_s}.
\end{align}
Both functionals can be defined in terms of a set of numerically convenient
one-particle orbitals $\{\phi_i\}$, from which the Slater
determinant wavefunction, $\iket{\Phi_s}$ for $\lambda=0$, is
constructed. These orbitals are defined to be unoccupied, occupied
singly or in spin-pairs, giving occupation factors $f_i\in\{0,1,2\}$.
Thus, e.g., we can write
$T_s=\sum_if_i\ibraketop{\phi_i}{\th}{\phi_i}$ for the KS kinetic energy
and $n=\ibraketop{\Phi_s}{\nh}{\Phi_s}=\sum_if_i|\phi_i|^2
\equiv\ibraketop{\Psi^{n,1}}{\nh}{\Psi^{n,1}}$ for the density.
The orbitals obey the Kohn-Sham equation
\begin{align}
  \big\{ \th + v_s[n](\vr) \big\} \phi_i[n](\vr)
  =&\epsilon_i[n]\phi_i[n](\vr).
\end{align}
Here $\th=-\half\nabla^2$ and $v_s[n]\equiv v^0[n]$ is the
single-particle multiplicative Kohn-Sham
potential, which is the fictitious effective potential experienced by
the orbitals.

The Kohn-Sham formulation of DFT therefore transforms a difficult
many-electron problem into a simpler non-interacting one. The
remaining complexity is bundled into a correlation term
$E_{\crm}[n]=F^1[n]-T_s[n]-E_{\Hx}[n]$ which is also a functional of the
density $n$}. $E_{\crm}$ is highly non-trivial in general, but can be
usefully approximated -- typically, but not always, in combination
with the exchange part $E_{\xrm}[n]$ of $E_{\Hx}[n]$ (as $E_{\xc}[n]$)
to allow for error cancellation. Many useful approximations for
$E_{\xc}$ exist that allow DFT to be used cheaply in a predictive
fashion (see, e.g.,
\rcites{Perdew2001,Kummel2008,Burke2012,Becke2014-Perspective,%
Jones2015-Perspective}).
When the correlation component is set to zero but the other quantities
are evaluated exactly one ends up with the ``exact exchange''
approximation.

\subsection{Ensemble density functional theory}

DFT was originally conceived as a theory of pure-states and
in its original form provides direct access only to properties
of the ground state, notably its electron density and energy.
DFT was later generalized to the case of
ensembles\cite{Theophilou1979,Valone1980},
which can be broadly categorized into three forms: First, there are
ensemble of states with different numbers of electrons in each
state;\cite{Perdew1982} Second, ensembles may be required to deal with
degenerate ground states\cite{Levy1982};
and finally, Gross, Oliveira and Kohn (GOK)
ensembles\cite{Gross1988-1,Gross1988-2,Oliveira1988-3}
extend density functional theory to statistical ensembles of
eigenstates.

Specifically, GOK ensemble DFT (EDFT) replaces a single groundstate
wavefunction by a density matrix
\begin{align}
  \Gammah_{\FW}=&\sum_{\FE}w_{\FE}\iout{\Psi_{\FE}},
  &
  \sum_{\FE} w_{\FE}=1,
  \label{eqn:GammaDef}
\end{align}
where $\ibraket{\Psi_{\FE}}{\Psi_{\FE'}}=\delta_{\FE\FE'}$, and
where the set of positive weights $\FW\equiv\{w_{\FE}\}$ obeys
certain constraints discussed below.
Following a similar sequence of steps to Eq.~\eqref{eqn:Evpure}, the
ensemble energy can be calculated through,
\begin{align}
  \fancy{E}[v;\FW]=&\min_n\bigg\{ \F^1[n;\FW]
  + \int n(\vr)v(\vr)d\vr \bigg\}
  \nonumber\\
  \equiv&\sum_{\FE}w_{\FE}E_{\FE}[v].
  \label{eqn:EvW}
\end{align}
where the minimization is performed over the \emph{statistically
  averaged} density
$n=\sum_{\FE}w_{\FE}\ibraketop{\Psi_{\FE}}{\nh}{\Psi_{\FE}}$,
and where $E_{\FE}[v]$ are the low lying eigenvalues of the
many-electron Hamiltonian $\Hh_v$.

One can then invoke the ensemble version of $F^{\lambda}[n]$,
\begin{align}
  \F^{\lambda}[n;\FW]=&\min_{\Gammah_{\FW}\to n}
  \tr[\Gammah_{\FW}(\Th+\lambda\Wh)]
  \label{eqn:Flambda}
\end{align}
which is subject, as above, to constrained minimization such that
$\tr[\Gamma_{\FW}\nh]\equiv\sum_{\FE}w_{\FE}
\ibraketop{\Psi_{\FE}}{\nh}{\Psi_{\FE}}=n(\vr)$, and defined for given
``well-behaved'' sets of fixed weights $\FW=\{w_{\FE}\}$.
Thus, $\fancy{E}$ now equals a statistical average of the lowest
lying energy eigenvalues $E_{\FE}[v]$ of
$\Hh_v=\Th+\Wh+\int \nh(\vr)v(\vr) d\vr$ for
weights $\FW=\{w_{\FE}\}$ obeying $\sum w_{\FE}=1$,
$0\leq w_{\FE}\leq 1$,
$w_{\FE}\geq w_{\FE}$ for $E_{\FE}\leq E_{\FE'}$ and other
conditions discussed in detail in the original GOK
articles{\cite{Gross1988-1,Gross1988-2,Oliveira1988-3}} and
in more recent work.\cite{Gould2017-Limits}

As above for the pure state, we can implicitly define a density
matrix
$\Gammah_{\FW}^{n,\lambda}\equiv \sum_{\FE}w_{\FE}
\iout{\Psinl_{\FE}}$
using $\tr[\Gammah_{\FW}^{n,\lambda}(\Th+\lambda\Wh)]
=\F^{\lambda}[n;\FW]$, i.e., $\Gammah_{\FW}^{n,\lambda}$ is any density
matrix that minimizes the trace which, in many cases, will not be
unique. Similarly, we can extend the idea of an \emph{ensemble}
$v$ representable
density\cite{Ayers2006-Axiomatic} to one for which the eigenstates
$\iket{\Psinl_{\FE}}$ in $\Gammah_{\FE}^{n,\lambda}$ obey
$[\Th+\lambda\Wh+\vh^{\lambda}-E^{n,\lambda}_{\FE}]\iket{\Psinl_{\FE}}=0$
with $v^1=v$ and, analogously to the pure ground state case,
$v_s[n,\FW]\equiv v^0$. The wavefunctions
$\iket{\Phi_{s,\FE}}\equiv\iket{\Psi_{\FE}^{n,0}}$ can then be written as a
set of orthogonal Slater determinants. Pure-state DFT, per
Eq.~\eqref{eqn:Evpure}, is the special case $w_{0}=1$ and $w_{\FE>0}=0$.

Thus, DFT can be generalized to include an ensemble like that of 
\eqref{eqn:GammaDef}, formed using a fixed set of
ensemble weights $\FW=\{w_{\FE}\}$,
which, as before, can be written in terms of a set of occupied KS
orbitals obeying
\begin{align}
  \big\{ \th + v_s[n;\FW] \big\}\phi_i[n;\FW]
  =&\epsilon_i[n;\FW]\phi_i[n;\FW],
\end{align}
where
\begin{align}
  v_s[n;\FW](\vr)\equiv&v(\vr) + v_{\Hxc}[n;\FW](\vr),
\end{align}
is the ensemble Kohn-Sham potential.
Here the one-body system depends on $n=\sum_if_i|\phi_i|^2$, as above. A key difference,
however, is that we must consider also the set of weights $\FW$ --
\emph{each unique set of weights defines a unique functional in a
  rigorous fashion}. This generalization away from a pure ground state
allows the Kohn-Sham occupation factors $f_i[n,\FW]\in[0,2]$ to take
on non-integer values in a rigorous fashion. Related discussion on the
topic of non-integer ensembles can be found in \rcite{Li2017}.

One can now ensemble-generalize other functionals.
The non-interacting kinetic energy functional, $\Ts[n;\FW]$ is
readily given by
\begin{align}
  \Ts[n;\FW]\equiv \F^0[n;\FW]\equiv
  \sum_if_i\ibraketop{\phi_i}{\th}{\phi_i}\;.
\end{align}
Given the density $n(\vr)$ and set of fixed ensemble weights
$\FW=\{w_{\FE}\}$, there also exists a unique
Hartree-exchange energy functional, given by\cite{Gould2017-Limits} 
\begin{align}\label{EEHx}
  \EHx[n;\FW]=&\lim_{\lambda\to 0^+}
  \frac{\F^{\lambda}[n;\FW]-\Ts[n;\FW]}{\lambda} \nonumber
  \\
  \equiv&\sum_{\FE}w_{\FE}\Lambda_{\Hx,\FE}[n;\FW].
\end{align}
Thus, the Hartree-exchange functional, $\EHx[n;\FW]$ can be defined
even though $\Gammah_{\FW}^{n,\lambda=0}$ is not necessarily
unique. Eq.~\eqref{EEHx} involves a
set of unique Hx energy functionals, $\Lambda_{\Hx,\FE}[n]$, one for
each weight $w_{\FE}$, which are ``block eigenvalues'' of an
interaction matrix $\mat{W}=W_{\FE\FE'}
=\ibraketop{\Phi_{s,\FE}}{\Wh}{\Phi_{s,\FE'}}$, involving only the set
of Kohn-Sham non-interacting Slater determinant states
$\iket{\Phi_{s,\FE}}$ included in the non-interacting ensemble. This
means that $\EHx$ is a functional of the (partially) occupied orbitals
only.  It can be shown\cite{Gould2017-Limits} that the energy
functionals $\Lambda_{\Hx,\FE}$ naturally allow the
overall functional to directly adapt to
fundamental spin symmetries without any external inputs or
assumptions, even when multi-reference physics is
required. The above definition reduces to the combined
Hartree-exchange proposed earlier by Nagy~\cite{Nagy2005} and to the
SEHX expression~\cite{Yang2014}
in certain special cases, including the one presented here.
Work by Filatov~\cite{Filatov2015,Filatov2016} uses similar
principles to those espoused in Ref.~\cite{Gould2017-Limits} to show
how EDFT can help with approximating strong correlations, for both
ground and excited states.

In the ``ensemble exact exchange'' (EEXX) approximation, 
$\Ts[n;\FW]$ and $\EHx[n;\FW]$ are evaluated exactly but
correlation (via ensemble-generalized
$\Ec[n;\FW]=\F^1[n;\FW]-\Ts[n;\FW]-\EHx[n;\FW]$)
is neglected. EEXX calculations can yield good results in small
atoms,\cite{Yang2014,Pribram-Jones2014,Yang2017-EDFT} even 
for excitations that are very difficult for approximations to
time-dependent Kohn-Sham theory. EEXX can be calculated in two ways:
it can be obtained as a functional of the exact density, using the
exact orbitals, which is the course we pursue in this work to avoid
density-driven errors\cite{Kim2013}.
More commonly, it is performed using orbitals obtained self-consistently
through an optimized effective potential approach.\cite{OEP1,OEP2}
Details of $\EHx$ that are relevant to the cases considered in the
remainder of this manuscript are discussed in
greater detail in Appendix~\ref{app:Lambda}.

\subsection{A numerically solvable model of CT excitations}

We choose a simple model diatom system  possessing two electrons in a
one-dimensional and (controllably) asymmetric diatomic molecule. We
define,
\begin{align}
  \Hh=&\Th + \Wh + \vh,
  \label{eqn:Ham}
\end{align}
where the kinetic energy operator is
$\Th=\th+\th'$ with $\th=-\half \frac{d^2}{dx^2}$, the external potential
operator is $\vh=\int dx \nh(x)v(x)$,
and the interaction operator is
$\Wh=\int \frac{dx dx'}{2} \nh_2(x,x')U(x-x')$,
where $\nh_2(x,x')=\nh(x)\nh(x')-\delta(x-x')\nh(x)$.
Here we employ a soft-Coulomb potential,
$U(z)=(\frac14+z^2)^{-\half}$, for Coulomb interactions. For
the external potential we use
\begin{align}
  v(x)=&-U\big(x+R/2\big)\nonumber\\&~~~
    - \big[ U\big(x-R/2\big) + \mu_S e^{-(x-R/2)^2} \big].
\end{align}
Here $R$ is the bond length between the left atom lying at $-R/2$ and
right atom at $+R/2$. The term $\mu_S$ changes the well depth
on the right atom, with larger $\mu_S$ making the well deeper.

By varying $\mu_S$ we are able to change the form of the
ground state in the dissociation limit, $R\to\infty$. For $\mu_S=0$, symmetry ensures
that both the left and right atoms have one electron each By contrast, for $\mu_S=2.0$ the
dissociation limit leads to two electrons on the right atom,
and none on the left, with the change in asymptotic behavior occurring
for $\mu_S\approx 1.4$.
Numerically, we find that for $0\leq \mu_S\leq 2$ the triplet state
always involves one electron on each of the two nuclei, meaning that
for sufficiently large $R$ and $\mu_S$, the lowest energy excitation
involves transferring charge from the right atom to the left, as in
Figure~\ref{fig:Fig0}. Thus we have a numerically solvable model which 
contains the key physics we wish to study, namely charge
transfer excitations.

We define the ground state as $\iket{\gs}\equiv
\iket{\Psi^{n,1}_0}$. For reasons of pedagogical simplicity, here we
focus on the lowest energy singlet-triplet transition and define the
lowest triplet excited state, $\iket{\ts}\equiv \iket{\Psi^{n,1}_1}$
(singlet excitations are discussed in Section \ref{subsec_singlet}
below). If we set  $w_0=1-p$ and $w_1=p$ we can define an ensemble
$\Gammah^{n,1}=(1-p)\iout{\Psi^{n,1}_0} + p\iout{\Psi^{n,1}_1}
=(1-p)\iout{\gs}+p\iout{\ts}$ that
is equivalent  to having a probability $p$ of being in the three-fold
degenerate lowest excited state%
\footnote{We work here entirely in a spin-unpolarized formalism, in
  which potentials do not depend on spin and the orbitals can be
  separated into a spin-independent spatial part and an explicit spin
  part. In this formalism it is a natural consequence of EDFT that one
  can arbitrarly choose any of the three triplet states with no 
  change in results.}
and a probability $(1-p)$ of being in the ground state.
We can then rewrite Eq.~\eqref{eqn:EvW} as
\begin{align}
  \E[v,p]
  =&\F^1[\np,p] + \int \np(x) v(x) dx, \nonumber
  \\
  =&w_{\gs}E_{\gs} + w_{\ts}E_{\ts}=E_{\gs} + p[E_{\ts}-E_{\gs}]
\end{align}
where $\np=n_{\gs}+p[n_{\ts}-n_{\gs}]$ is the density of the ensemble
system [parametrized using $p$, as indicated by the superscript
  $(p)$] with external potential $v$.
Thus, we obtain an energy that depends linearly on the excitation
energy $E_{\ts}-E_{\gs}$, which allows us to use Eq.~\eqref{eqn:EvW}
to calculate energy differences by varying $p$.
Here and henceforth we restrict the
set of weights $\FW$ to provide such an admixture of the ground- and
excited states only, i.e., we set $w_0=1-p$, $w_1=p$ and
$w_{\FE>2}=0$ as above. We can therefore adopt a short-hand
notation, $\E^{(p)}\equiv \E[n=\np,\FW=\{1-p,p\}]$.

We can determine the exact eigenstates of our model Hamiltonian
\eqref{eqn:Ham} using simple
numerics implemented in Python with NumPy and SciPy.
This lets us calculate properties, such as energies, energy
differences, and densities for the true ensemble $\Gammah^{n,1}$.
From the exact results, we can then use density inversion techniques
for EDFT\cite{Gould2014-KS} to obtain the non-interacting KS reference
system. This involves finding a multiplicative potential,
  $v_s^{(p)}$, that yields single-particle orbital solutions of
\begin{align}
  &\big\{ \th + v_s^{(p)}(\vr) \big\}\phi_i^{(p)}(\vr)
  =\epsilon_i^{(p)}\phi_i^{(p)}(\vr),
  \label{eqn:hKS}
\end{align}
such that they correctly reproduce the target density, i.e.,
\begin{align}
  \np=&(1-p)n_{\gs} + p n_{\ts}
  =(1-p)n^{(p)}_{s,\gs} + p n^{(p)}_{t,\ts}
  \nonumber\\
  =& (2-p)|\phi_0^{(p)}|^2+p|\phi_1^{(p)}|^2,
  \label{eqn:nEquiv}
\end{align}
where the last line uses the relations
$n_{s,\gs}(\vr)=2|\phi_0(\vr)|^2$ and
$n_{s,\ts}(\vr)=|\phi_0(\vr)|^2+|\phi_1(\vr)|^2$,
which connect between the densities of the Kohn-Sham ensemble members and the
Kohn-Sham orbitals.
When \eqref{eqn:hKS} and \eqref{eqn:nEquiv} are simultaneously
satisfied, $v_s^{(p)}\equiv v+v_{\Hxc}^{(p)}$ is the exact Kohn-Sham
potential and, thus, $v_{\Hxc}^{(p)}$ is the exact Hartree-exchange-
correlation potential.

Importantly, and unlike previous work on excited
states using unrestricted Hartree-Fock theory,\cite{Barca2014-HF}
we adopt a spin-restricted framework, i.e., spinors have an equal
spatial component for either up $\up$ or down $\down$ single-particle
states, thereby avoiding any symmetry breaking. 
Thus our ensembles account for eigenstates of both
$\hat{S}^2$ and $\hat{S}_z$. Similarly we preserve the mirror symmetry
of H$_2$ ($\mu_S=0$). We thus preserve as many exact conditions as we can.

The exact KS orbitals allow us to calculate all the reference data
for the analyses reported in the next section and compare to
approximate KS data. For our tests we
make the Kohn-Sham ensemble exact exchange ($\EEXX$) approximation,
\begin{align}
\F[n,\FW]\approx \Ts[n,\FW]+\EHx[n,\FW],
\end{align}
as an extension of its ground state counterpart, i.e., our only
approximation is to set $\Ec[n,\FW]\equiv 0$.
Thus, for arbitrary $p$ and exact orbitals $\phi_i^{(p)}$, we have
\begin{align}
  \E_{\EEXX}^{(p)}=& \Ts^{(p)} + \EHx^{(p)}
  + \int \np v dx
  \\
  \equiv& (1-p)\{T_{s,\gs}^{(p)}+\Lambda_{\Hx,\gs}^{(p)}\}
  \nonumber\\&
  + p\{T_{s,\ts}^{(p)}+\Lambda_{\Hx,\ts}^{(p)}\}
  + \int \np v dx.
  \label{eqn:EWKS}
\end{align}
The kinetic and interaction energy terms
have implicit (via the orbitals) and explicit $p$ dependencies.
The kinetic energy terms for the states are
\begin{align*}
  T_{s,\gs}^{(p)}=&2t_0^{(p)},
  ~~~~~~~~~~~~~~
  T_{s,\ts}^{(p)}=t_0^{(p)}+t_1^{(p)},
\end{align*}
where $t_i=\int \phi_i(x)\th\phi_i(x)dx$ and all orbitals
$\phi_i$ are real. The interaction energy terms,
\begin{align}
  \Lambda_{\Hx,\gs}^{(p)}=&
  \int \frac{dxdx'}{2}U(x-x')2\phi_0^{(p)}(x)^2\phi_0^{(p)}(x')^2
  \label{eqn:Lambdags}
  \\
  \Lambda_{\Hx,\ts}^{(p)}=&\int \frac{dxdx'}{2}U(x-x')
  \nonumber\\&\times
         [\phi_0^{(p)}(x)\phi_1^{(p)}(x')-\phi_1^{(p)}(x)\phi_0^{(p)}(x')]^2
  \label{eqn:Lambdats}
\end{align}
are defined according to the underlying symmetries of the singlet
ground- and triplet excited states -- which follows directly from the
definition of $\EHx$\cite{Gould2017-Limits} (see
Appendix~\ref{app:Lambda} for details).

\section{Results}

Having established the theory and model systems, we now 
report the results of several tests that examine the successes and
limitations of the proposed EDFT approach. 

\subsection{Triplet states}
\label{subsec_triplet}
First, we establish that
exact EDFT does indeed capture the nature of charge transfer
excitations. To this end, we now consider the density components that
comprise the statistical ensemble, in order to examine the ability of the
approach to ``move'' charge during excitations (as illustrated in
Figure~\ref{fig:Fig0}, where one electron is moved from the right atom
to the left one under excitation).

We determine charge densities for the ground and triplet states in two
different ways. First, we define $n_{\gs}=\ibraketop{\gs}{\nh}{\gs}$
and $n_{\ts}=\ibraketop{\ts}{\nh}{\ts}$ to be the true electron
densities of the ground state and triplet wavefunctions, respectively.
Next, $n_{s,\gs}^{(p)}(x)=2\phi_0^{(p)}(x)^2$ and
$n_{s,\ts}^{(p)}(x)=\phi_0^{(p)}(x)^2+\phi_1^{(p)}(x)^2$
are the densities of the corresponding Kohn-Sham states
$\iket{\Phi_{s,\gs/\ts}}$, obtained by minimizing $\Ts=\F^0$ subject
to the constraints.
Note that generally $n_{\gs}\neq n_{s,\gs}^{(p)}$ (except for $p=0$) and
$n_{\ts}\neq n_{s,\ts}^{(p)}$, i.e., the KS ground-state
and triplet state densities do not need to be the same as the exact
ones even in exact EDFT. Only the statistical average of the KS
density must equal that of the density of the interacting system,
i.e., $\np=(1-p)n_{\gs}+p n_{\ts}=(1-p)n_{s,\gs}^{(p)}+p n_{s,\ts}^{(p)}
=\np_{s}$ [cf. Eq.\ \eqref{eqn:nEquiv}
and see Appendix~\ref{app:Densities} for further discussion].

\begin{figure}[t]
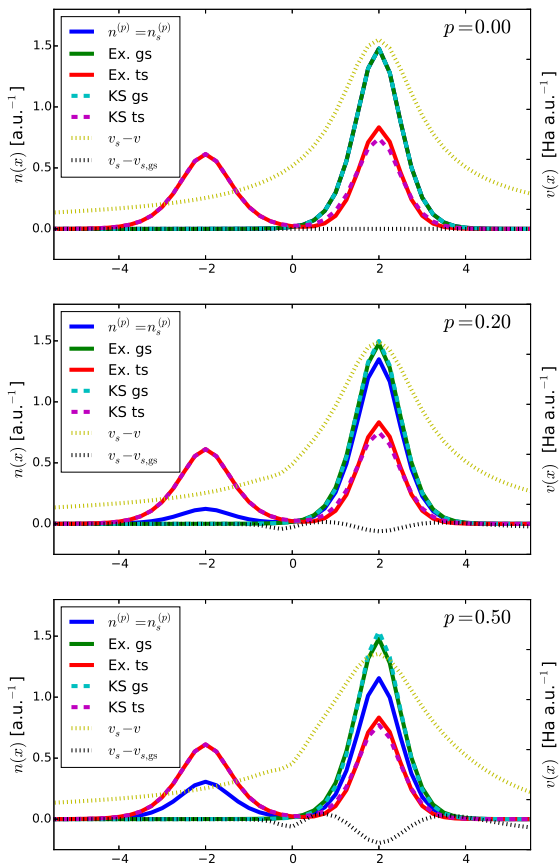

  \includegraphics[width=0.9\linewidth]{{{Dens_04.0_1.00_1.50_0.00}}}\\
  \includegraphics[width=0.9\linewidth]{{{Dens_04.0_1.00_1.50_0.20}}}\\
  \includegraphics[width=0.9\linewidth]{{{Dens_04.0_1.00_1.50_0.50}}}
  \caption{Exact ($n_{\gs/\ts}$, solid lines) and Kohn-Sham
    ($n^{(p)}_{s,\gs/\ts}$, dashed lines) densities of the ground- and first
    excited states with $R=4$ and $\mu_S=2.0$, calculated from the
    interacting and non-interacting wave-functions, respectively.
    Top: $p=0$, middle: $p=0.2$, bottom: $p=0.5$.
    In all cases the KS states are found to be good representations of the exact densities 
    despite not being under any ``formal'' obligation to be so.
    Also shown (in dotted lines) are the Hxc potential, $v_{\Hxc}^{(p)}$, and the ensemble potential difference,
    $v_s^{(p)}-v_s^{(0)}$.
  \label{fig:DensGood}}
\end{figure}

Figure~\ref{fig:DensGood} shows interacting-system (solid lines)
and exact Kohn-Sham (dashed lines) densities, as obtained from the
above-described inversion process, for the case
of $R=4$ and $\mu_S=2$ with $p=0$, $p=0.2$, and $p=0.5$.
For all $p$, the ground-state and triplet densities of the real and
KS states, while indeed not equal, are clearly similar, demonstrating
a genuine ability of the EDFT to transfer charge spatially.
This is a non-trivial result as the individual KS densities are
only constrained by their ensemble average. Thus, e.g.. in the case
$p=0.5$ the KS system could have had 1.5~electrons on the right
atom and 0.5~electrons on the left in both the ground and triplet
states, as in the total density. That the individual KS densities
resemble their exact counterparts, with 2~electrons in the right
atom for the ground state and 1~electron on each atom for the
triplet state, is therefore a success of KS EDFT.
Filatov~\emph{et al} have similarly shown that approximations
to EDFT can describe transfer of charge in excitations of the
4-(N,NDimethyl-amino)benzonitrile (DMABN)
chromophore, albeit without direct comparison to the densities
of the exact transitions.\cite{Filatov2015}

The plots in Figure~\ref{fig:DensGood} also include (as dotted lines)
the exact Hartree-exchange-correlation potential
$v_{\Hxc}^{(p)}=v_s^{(p)}-v$, as well as the difference between the KS
potential obtained at finite $p$ with that obtained for the pure
ground state, i.e., $v_s^{(p)}-v_s^{(0)}$.  Importantly, it is
well-known that in open electron-number ensemble systems, the addition
of a small amount of additional charge can lead to
difficult-to-approximate step
features.\cite{Perdew1982,Gould2014-KS,Karolewski2011,Kraisler2015}
The exact potentials plotted in Figure~\ref{fig:DensGood} exhibit no
such features. This highlights a potential advantage of EDFT over
alternative approaches, in that the ensemble correction to the KS
system may lend itself to future approximations involving semi-local
functionals that cannot produce step-like features.

Having established the validity and potential usefulness of the EDFT approach, we turn to examining energy differences in charge transfer states. 
We have already established that $\E^{(p)}=E_{\gs} + p(E_{\ts}-E_{\gs})$, where
$E_{\gs}$ and $E_{\ts}$ are defined for a given $v$ that is determined by
$R$ and $\mu_S$, with the pure ground state, $E_{\gs}=\E^{(0)}$,
obtained for $p=0$.
For the exact functional, then, the energy is a straight line in $p$,
without any implicit dependence on $p$, yielding
\begin{align}
  \Omega\equiv E_{\ts}-E_{\gs}=\frac{\E^{(p)}-\E^{(0)}}{p}
  =\frac{\partial \E^{(p)}}{\partial p}
  \label{eqn:Omega}
\end{align}
for the exact excitation energy (optical gap) from the ground to
triplet state. We can compare these exact results to approximate ones
obtained using the exact-exchange expression [Eq.~\eqref{eqn:EWKS}],
where the correlation energy is neglected. This means that the
approximate expressions
\begin{align}
  \Omega_{\EEXX}^{(p)}\equiv& \frac{\E_{\EEXX}^{(p)}-\E_{\EEXX}^{(0)}}{p},
  \label{eqn:OmegaLine}
\end{align}
or
\begin{align}
  {\Omega'}_{\EEXX}^{(p)}\equiv&
  \frac{\partial \E_{\EEXX}^{(p)}}{\partial p},
  \label{eqn:OmegaDerv}
\end{align}
are neither necessarily the same nor necessarily independent of $p$, due to
implicit dependencies on the orbitals.

\begin{figure}[t]
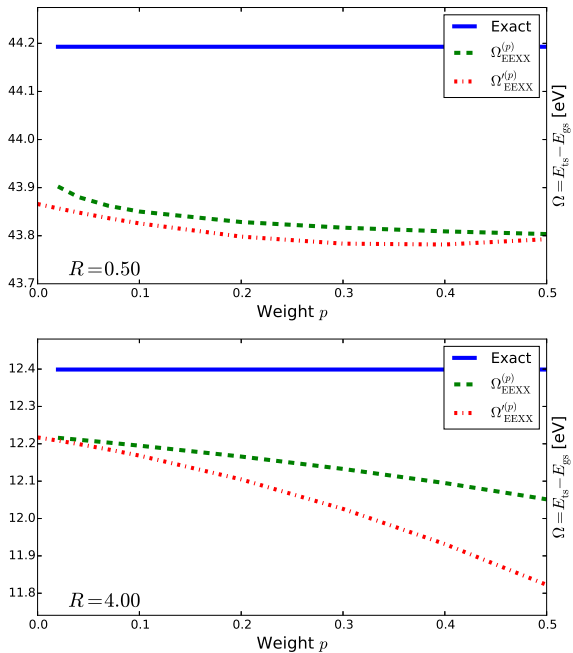

  \includegraphics[width=0.9\linewidth]{{{Delta_00.5_1.00_1.50_All}}}\\
  \includegraphics[width=0.9\linewidth]{{{Delta_04.0_1.00_1.50_All}}}
  \caption{Exact energy gap (as obtained in both the many-electron and
    the exact Kohn-Sham system), compared with that obtained in the
    EEXX approximation calculated in two different ways, based on
    $\Omega_{\EEXX}$ and $\Omega'_{\EEXX}$
    [Eqs.~\eqref{eqn:OmegaLine} and \eqref{eqn:OmegaDerv}],
    with $R=0.5$ (top) and $R=4$ (bottom) and $\mu_S=2.0$, which
    defines a clear charge transfer excitation.
    The difference between $\Omega_{\EEXX}$ and $\Omega'_{\EEXX}$
    for $W\to 0$ for $R=0.5$ is due to numerical errors.
  \label{fig:DeltaGood}}
\end{figure}

The results of the exact calculations for $\Omega$, compared with
approximate ones obtained using both $\EEXX$ excitation expressions
given above, at different values of $p$, are given in
Figure~\ref{fig:DeltaGood}. We use $\mu_S=2$, which corresponds to a
charge transfer molecule, and study both $R=0.5$ and
$R=4$. Importantly, here and below the approximate results are not
obtained self-consistently, but rather from the approximate energy
expression based on the exact densities. This allows us to focus on
errors due to the approximate functional and eliminate errors due to
an approximate density.\cite{Kim2013} Figure~\ref{fig:DeltaGood} shows
that the approximate expressions yield results that are within a few
tenths of an eV of each other and in generally similar agreement with
exact results, with the non-derivative expression
\eqref{eqn:OmegaLine} yielding a curve that is somewhat flatter
and in better agreement with the exact value. This is quite
satisfactory, given that no correlation energy is included.

\begin{figure}[t]
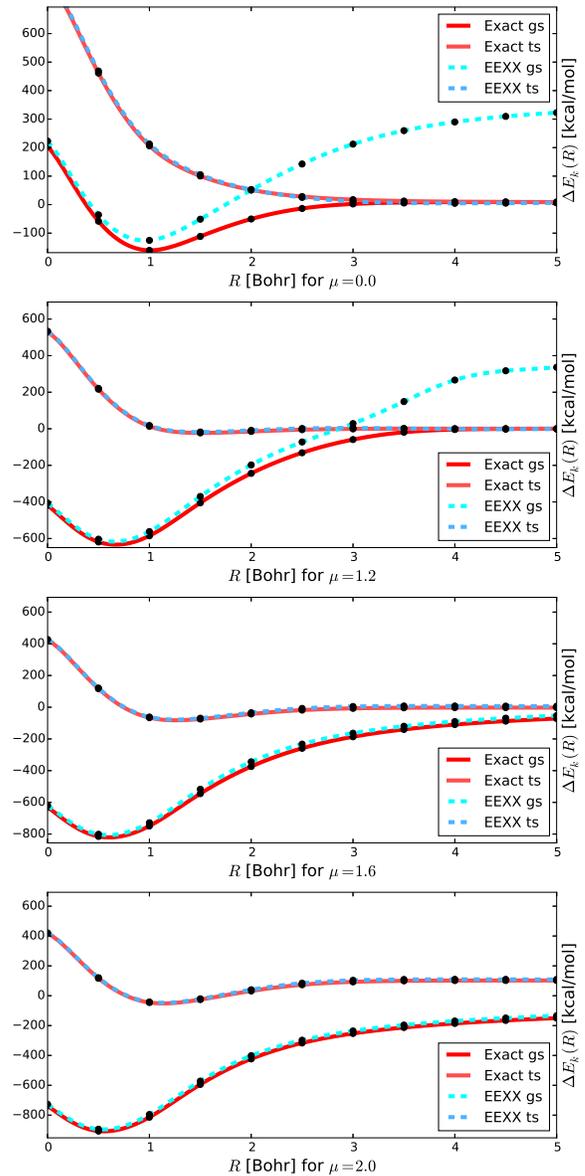

  \includegraphics[width=0.9\linewidth]{{{Energies_1.00_1.00}}}\\
  \includegraphics[width=0.9\linewidth]{{{Energies_1.00_1.30}}}\\
  \includegraphics[width=0.9\linewidth]{{{Energies_1.00_1.40}}}\\
  \includegraphics[width=0.9\linewidth]{{{Energies_1.00_1.50}}}\\
  \caption{
    Exact and Hartree-exchange energies dissociation curves for the
    ground state and triplet state for $\mu_S=0$ (top), $1.2$ (second),
    $1.6$ (third) and $2$ (bottom). EEXX energies are obtained using
    Eq.~\eqref{Diss_t}.
    Remarkably, in all cases Hartree-exchange energies are excellent
    approximations to the triplet energy, even when strong static
    correlation results in very poor ground state energies that can
    even be higher in energy than the excited state.
  \label{fig:Curves}}
\end{figure}

Finally, we consider the ability of EEXX to reproduce dissociation
curves for either the ground state or the triplet state, defined by
$\Delta E_{\gs/\ts}(R)=E_{\gs/\ts}(R)-E_{\gs}(R\to\infty)+U(R)$, where
the penultimate term is the ground-state energy at the full
dissociation limit and the final term is the inter-nuclear repulsion
energy. A comparison between EEXX and exact EDFT is given in
Figure~\ref{fig:Curves}, where results are shown for two
strongly-correlated dimers ($\mu_S=0$ and $1.2$) and two
charge-transfer dimers ($\mu_S=1.6$ and $2$). The triplet-state EEXX
results were obtained via the relation
\begin{align}\label{Diss_t}
  E_{\EEXX,\ts}(R)\equiv & E_{\EEXX,\gs}(R) + \Omega^{(0.5)}_{\EEXX}(R),
\end{align}
where $ \Omega^{(0.5)}_{\EEXX}(R)=
2[ \E_{\EEXX}^{(0.5)}(R)-\E_{\EEXX}^{(0)}(R) ]$,
i.e., the excitation energy is evaluated at the maximal mixing point,
$p=0.5$, using a difference formula.

Clearly, for the charge-transfer dimers ground-state dissociation
curves are well-reproduced by EEXX. However, for the
strongly-correlated dimers the ground-state dissociation curves are
very poorly-reproduced, to the point that the energies become
\emph{greater} than the excited state in the dissociation limit, which
means that the predicted Kohn-Sham excitation energy is
\emph{negative}, at the Hx level.  The failure of a zero-correlation
expression in the strong correlation limit is not at all surprising in
itself. What may seem counterintuitive, however, is the negative
excitation energy. This is because DFT, even in GOK ensemble form, is
a theory of lowest energy states and thus one expects that other
states should be energy-ordered accordingly under any DFT
approximation.  Nevertheless, this result is perfectly in line with
the theory, because the universal functional $\F[n,\FW]$ is defined
for a \emph{given choice} of $\FW$ and $n$. Thus, when we choose
$p=0$ and $p=0.5$ we are using \emph{different density functionals}
and there is no issue with ordering when comparing energies as we do
here.

Remarkably, triplet-energy dissociation curves for the
charge-transfer dimers are well-reproduced at all $R$ and for
\emph{all} dimers, including the most correlated H$_2$ molecule
($\mu_S=0$), despite a ground-state that is a very poor approximation
for the strongly-correlated true ground state. \cite{Dunlap1983}
Indeed, a higher-quality triplet state, compared to the ground state,
was reported previously using hybrid functional theory in the context
of triplet instabilities.\cite{Peach2011}

\subsection{Singlet states}
\label{subsec_singlet}

As mentioned in our introduction of the model system, we have focused on the
the lowest energy singlet-triplet transition for reasons of pedagogical simplicity. 
However, this poses significant limitations. First, the singlet-triplet transition is ``optically dark'' and therefore of less practical interest; Second, it is actually amenable to analysis using conventional ground state DFT, if appropriate spin-symmetry restrictions are imposed. 
Therefore, in this section we discuss a more general ensemble that includes
contributions from the lowest-lying excited singlet state and use it to study the physically important, and more difficult to reproduce, singlet CT excitation.

Consider a GOK ensemble with a mixture of $p\leq\frac12$ triplet and
singlet excited states, of which a fraction $\beta\leq\frac14$
are in the singlet state.
(the upper bounds come from the general condition on GOK ensemble
weights that $w_{\FE}\geq w_{\FE'}$ when $E_{\FE}\leq E_{\FE'}$)
Therefore, we have
\begin{align}
  \Gammah=(1-p)\iout{\gs} + p(1-\beta)\iout{\ts}
  + p\beta\iout{\ss}
\end{align}
where $\iket{ss}$ is the first excited singlet state. This yields 
\begin{align}
  \E^{(p;\beta)}=E_{\gs}+p[ (E_{\ts}-E_{\gs})
    + \beta(E_{\ss}-E_{\ts}) ]
\end{align}
(note, $\E^{(p;0)}\equiv\E^{(p)}$) and
\begin{align}
  n^{(p;\beta)}=&(1-p)n_{\gs} + p[(1-\beta)n_{\ts} + \beta n_{\ss}]
  \nonumber\\&
  =(1-p)n^{(p;\beta)}_{s,\gs} + pn^{(p;\beta)}_{s,\ts}
  \nonumber\\&
  =(2-p)|\phi_0^{(p;\beta)}|^2+p|\phi_1^{(p;\beta)}|^2\;,
\end{align}
for the energy and density, respectively.
Here we used $n_{s,\ts}=n_{s,\ss}=|\phi_0|^2+|\phi_1|^2$, which
follows directly from the KS ensemble minimization. The kinetic energy
$\Ts^{(p;\beta)}=(2-p)t_0^{(p;\beta)} + pt_1^{(p;\beta)}$ takes the
same form as for the triplet state (but not the same value, as the
Kohn-Sham orbitals for this ensemble are different) and so do the
lowest two Hartree-exchange block eigenvalues
[given by Eqs.~\eqref{eqn:Lambdags} and \eqref{eqn:Lambdats}].
The singlet state has the block eigenvalue
\begin{align}
  \Lambda_{\Hx,\ss}=&\int \frac{dxdx'}{2}U(x-x')
         [\phi_0(x)\phi_1(x')+\phi_1(x)\phi_0(x')]^2
         \;,
  \label{eqn:Lambdass}
\end{align}
finally yielding the EEXX energy as
\begin{align}
  \E_{\EEXX}^{(p;\beta)}=&\Ts^{(p;\beta)}+(1-p)\Lambda_{\Hx,\gs}^{(p;\beta)}
  + p\Lambda_{\Hx,\ts}^{(p;\beta)}
  \nonumber\\&
  + p\beta[\Lambda_{\Hx,\ss}^{(p;\beta)}-\Lambda_{\Hx,\ts}^{(p;\beta)}]
  + \int n^{(p;\beta)} v dx.
\end{align}

\begin{figure}[t]
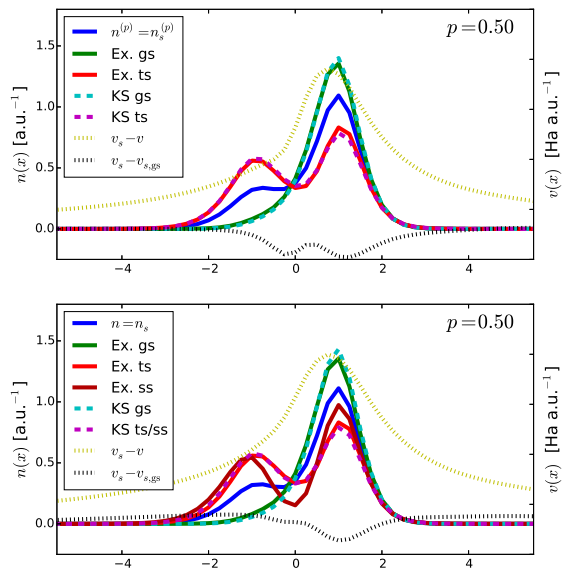

  \includegraphics[width=0.9\linewidth]{{{Dens_02.0_1.00_1.50_0.50}}}
  \includegraphics[width=0.9\linewidth]{{{Dens_02.0_1.00_1.50_0.50_S}}}
  \caption{
    Exact ($n_{\gs/\ts(/\ss)}$, solid lines) and Kohn-Sham
    ($n^{(p)}_{s,\gs/\ts(/\ss)}$, dashed lines) densities of the
    ground- and low-lying
    excited states with $R=2$ and $\mu_S=2.0$, calculated from the
    interacting and non-interacting wave-functions, respectively, for
    $p=0.5$ and $\beta=0$ (triplet excitation only, top) and 
    $\beta=0.25$ (singlet excitation included, bottom). 
    Also shown (in dotted lines) are the Hxc potential,
    $v_{\Hxc}^{(p)/(p,0.25)}$, and the ensemble potential difference,
    $v_s^{(p)/(p,0.25)}-v_s^{(0)}$.
  \label{fig:DensS}}
\end{figure}

With this reasonably straightforward generalization of the pedagogical
triplet case, we can now test the suitability of our approach
to singlet excitations.
To begin our analysis, we show in Figure~\ref{fig:DensS} the densities
of the exact ground-, triplet-, and singlet- states (solid lines),
and  their KS counterparts (dashed lines) for the difficult case of $R=2$ and
$\mu_S=2$. In this case, the singlet and triplet states possess qualitatively
different densities, which
must nevertheless still be accommodated by a single KS potential (for
the case of $R=4$, studied in Fig.~\ref{fig:DensGood} above, the
singlet/triplet densities are nearly
identical, as expected for a negligible singlet-triplet separation).
As before, the ground state
density is well-reproduced. The triplet-singlet average density
is also well-reproduced and is dominated by the contribution from the
triplet state, which is to be expected given its 75\% contribution.
The KS potential (dots) shows significant differences with respect to
that found in the previous sub-section (compare
Fig.~\ref{fig:DensGood}), reflecting the different ensemble
densities. Here the KS potential 
appears to have a small step-like feature on the right molecule,
although this may be a numerical artifact arising from the density inversion.
In any case, the step is still small compared to other features and
compared to the steps arising in the KS potential of conventional
DFT.

\begin{figure}[t]
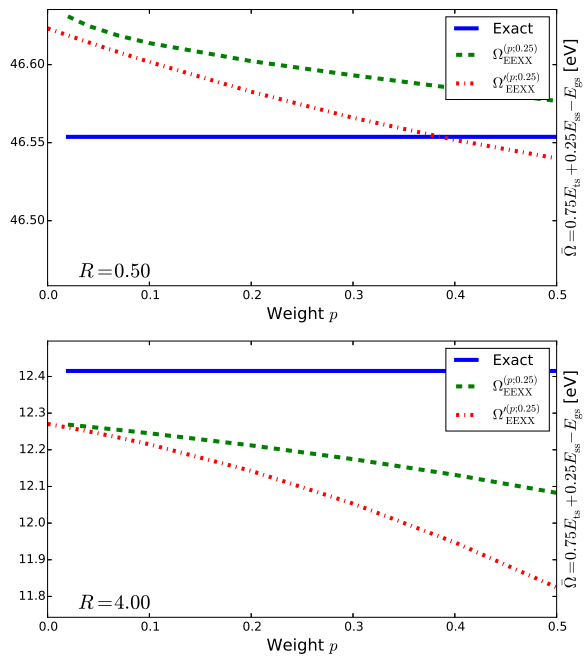

  \includegraphics[width=0.9\linewidth]{{{Delta_00.5_1.00_1.50_All_S}}}
  \includegraphics[width=0.9\linewidth]{{{Delta_04.0_1.00_1.50_All_S}}}
  \caption{
    Exact singlet-triplet averaged energy gap (as obtained in both the
    many-electron and the exact Kohn-Sham system),
    $\bar{\Omega}^{(0.25)}$, compared with that obtained in the two
    EEXX approximations, $\bar{\Omega}_{\EEXX}^{(p;0.25)}$ and
    $\bar{\Omega}'{}_{\EEXX}^{(p;0.25)}$, with $R=0.5$ (top) and $R=4$
    (bottom) and $\mu_S=2.0$, which defines a clear charge transfer
    excitation.
    \label{fig:DeltaS}}
\end{figure}

The singlet-triplet averaged gap, defined as 
\begin{align}
  \bar{\Omega}^{(\beta)}=(1-\beta)E_{\ts}+\beta E_{\ss} - E_{\gs}
  \equiv \Omega + \beta \Omega_{\st},
\end{align}
is shown in Figure~\ref{fig:DeltaS} both exactly and in the two EEXX approximations,
\begin{align}
  \bar{\Omega}_{\EEXX}^{(p;\beta)}
  =&\frac{\E_{\EEXX}^{(p;\beta)}-\E_{\EEXX}^{(0;\beta)}}{p},
  &
  \bar{\Omega}'{}_{\EEXX}^{(p;\beta)}
  =&\frac{\partial \E_{\EEXX}^{(p;\beta)}}{\partial p},
\end{align}
for $0\leq p\leq 0.5$.
Here $\Omega$ is the optical gap from Eq.\ \eqref{eqn:Omega} and
$\Omega_{\st}=E_{\ss}-E_{\ts}$ is the singlet-triplet splitting energy.
For $R=4$ and $\mu_S=2$ (bottom), the results are almost identical to
the ones given above, reflecting the fact that the singlet-triplet
splitting is very small. But for $R=0.5$ and $\mu_S=2$ (top), the
results are quite different, with the EEXX approximation
overestimating the singlet-triplet splitting
and thus compensating for some of the missing correlations that led to
under-prediction of the excitation energy in the pure triplet example.

\begin{figure}[t]
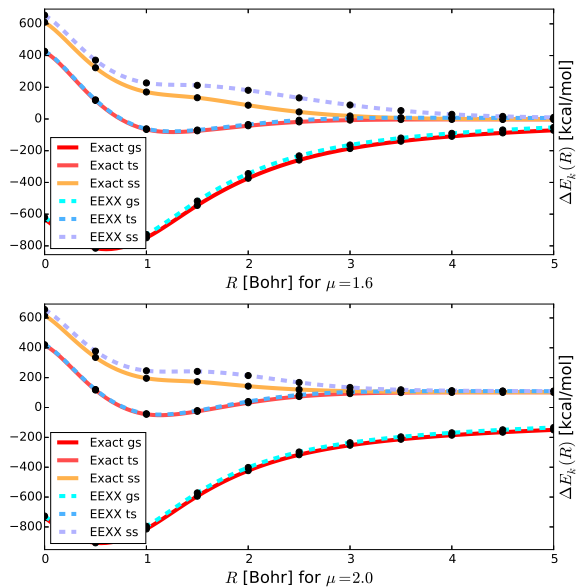

  \includegraphics[width=0.9\linewidth]{{{Energies_1.00_1.40_S}}}\\
  \includegraphics[width=0.9\linewidth]{{{Energies_1.00_1.50_S}}}\\
  \caption{
    Exact and Hartree-exchange energies dissociation curves for the
    ground state, singlet, and triplet state for $\mu_S=1.6$ (top) and
    $2.0$ (bottom). EEXX energies are obtained using Eq.~\eqref{Diss_s}.
    The agreement between exact and approximate singlet results is
    not as good as in the ground- and triplet states, but still has good
    semi-quantitative behavior.
  \label{fig:CurvesS}}
\end{figure}

Finally, Figure~\ref{fig:CurvesS} reproduces the energy curves for the
ground- and triplet- states already shown in Figure~\ref{fig:Curves}, but
includes also the first excited singlet state energy curve
$\Delta E_{\ss}(R)=\Delta E_{\ts}(R) + \Omega_{\st}(R)$
calculated exactly and at the EEXX level using
\begin{align}\label{Diss_s} 
    \Delta E_{\EEXX,\ss}(R)=\Delta E_{\EEXX,\ts}(R) + \Omega_{\EEXX,\st}(R)
  \end{align}
where $\Omega_{\EEXX,\st}(R)
=4[\bar{\Omega}^{(0.5,0.25)}_{\EEXX}-\Omega^{(0.5)}_{\EEXX}]$
is the EEXX singlet-triplet splitting energy calculated at
$p=0.5$ and $\beta=0.25$.

The excited singlet energy dissociation curve obtained with EEXX is
not as accurate as in the cases of the ground- and triplet states.
This is not surprising, as its energy is likely to have a greater
contribution from dynamical correlations which are unaccounted for in
EEXX. Nevertheless, the EEXX curve shows good semi-quantitative
agreement with the true curve, suggesting that one may devise
correlation approximations that can compensate for much of the
error. Dissociation curves for cases with stronger correlation
(such as $\mu_S=0,1.2$, not shown) are, as expected from the poor
singlet ground state in these cases, worse.

\section{Conclusion}

In this Article, we have shown that exact ensemble density functional
theory (EDFT), obtained through numerical inversion, can capture
charge transfer excitations without relying on time-dependent
calculations. In all cases, Kohn-Sham components of the ensemble
density were shown to possess a direct physical meaning, despite not
being constrained to achieve that.

Approximate excitation energies were obtained at the level of a
rigorously extended Hartree-exchange
approximation.\cite{Gould2017-Limits} Results for the triplet state
were shown to be good across an entire dissociation curve even when
the ground state is bad. For excited singlet state energies,
quantitative agreement was not as good as for the ground- and
triplet- states, likely owing to dynamic correlation effects. Still,
the transitions were well-predicted as long as strong correlations
were not present.

Importantly, the effective Kohn-Sham
potential needed to produce these results was found to lack a
difficult-to-approximate complex step structure that can appear in
other formalisms, at least when only triplets were considered.
A small step may be present in the difficult-to-reproduce
case of an excited singlet state with a density highly unlike that of the
corresponding triplet state; even then it is significantly
smaller in magnitude than other features of the potential.
This may indicate that the effective potential for ensembles is more
amenable to useful approximations for the difficult case of molecular
dissociation than the potentials in other density-based formulations.

Strictly speaking, the calculations presented here apply
to simplified, one-dimensional model systems. In particular, the role
played by differences between the densities and their non-interacting
KS counterparts warrants further consideration. Nevertheless, we
believe that these results are sufficiently fundamental to be
replicated in more realistic molecules, a case further
supported by recent approximate EDFT work.\cite{Filatov2015}
This work provides robust previously unavailable benchmarks and
provides an impetus for establishing EDFT correlation functionals that
will allow systematic improvements.

\acknowledgments

L.K. acknowledges support by the Israel Science Foundation.

\appendix
\section{The {$\Lambda_{\Hx}$} functionals}
\label{app:Lambda}

We summarize here the key features of $\Lambda_{\Hx}$ in the case of
the ground- and lowest lying excited state of ``typical'' systems
without spatial degeneracies.
The key to deriving these expressions is to recognize that
$\Lambda_{\Hx}[n;\FW]$ are eigenvalues of block sub-matrices of
$\ibraketop{\Psi_{\FE}}{\Wh}{\Psi_{\FE'}}$, taken over states with
identical densities and kinetic energies, and ordered from smallest to
largest within each block. Full details, and derivation, of the
minimization procedure used to derive the resulting ``block
eigenvalues'' can be found in the main article and supplementary
material of \rcite{Gould2017-Limits}.

In the case considered here, the KS ground state with $\phi_0$ doubly
occupied is non-degenerate, and therefore no other state shares its
density $n_{s,\gs}=2|\phi_0|^2$ or kinetic energy
$T_{s,\gs}=2t_0$. The first excited state is \emph{four-fold}
degenerate at the density/kinetic energy level, however, as the states
$\phi_0$ and $\phi_1$ can take on any combination of $\up$ and $\down$
spins in our spin-unpolarized formalism, while preserving
$n_{s,\ts}=n_{s,\ss}=|\phi_0|^2 + |\phi_1|^2$ and
$T_{s,\ts}=T_{s,\ss}=t_0+t_1$. Note that here these states are all
degenerate -- the triplet/singlet splitting is distinguished only
in the next step.

Because it is non-degenerate, we can calculate
$\Lambda_{\Hx,\gs}=\ibraketop{0\up,0\down}{\Wh}{0\up,0\down}$
directly for use in $\EHx$. But once triplet and singlet states are
involved we must find the eigenvalues of
\begin{align*}
  \mat{W}=&
  \begin{pmatrix}
    \ibraketop{\up\up}{\Wh}{\up\up} & 
    \ibraketop{\up\up}{\Wh}{\up\down} & 
    \ibraketop{\up\up}{\Wh}{\down\up} & 
    \ibraketop{\up\up}{\Wh}{\down\down}
    \\
    \ibraketop{\up\down}{\Wh}{\up\up} & 
    \ibraketop{\up\down}{\Wh}{\up\down} & 
    \ibraketop{\up\down}{\Wh}{\down\up} & 
    \ibraketop{\up\down}{\Wh}{\down\down}
    \\
    \ibraketop{\down\up}{\Wh}{\up\up} & 
    \ibraketop{\down\up}{\Wh}{\up\down} & 
    \ibraketop{\down\up}{\Wh}{\down\up} & 
    \ibraketop{\down\up}{\Wh}{\down\down}
    \\
    \ibraketop{\down\down}{\Wh}{\up\up} & 
    \ibraketop{\down\down}{\Wh}{\up\down} & 
    \ibraketop{\down\down}{\Wh}{\down\up} & 
    \ibraketop{\down\down}{\Wh}{\down\down}
  \end{pmatrix},
\end{align*}
where $\iket{\sigma\sigma'}$ is short-hand for 
$\iket{0\sigma,1\sigma'}$, to determine $\EHx$.
One can use the Slater-Condon rules to
eliminate many of the terms in $\mat{W}$, from which one finds the
three-fold degenerate lowest eigenvalue
$\Lambda_{\Hx,\ts}
=\ibraketop{0\up,1\up}{\Wh}{0\up,1\up} =
\ibraketop{0\down,1\down}{\Wh}{0\down,1\down} =
\frac{1}{\sqrt{2}}[
\ibraketop{0\up,1\down}{\Wh}{0\up,1\down} -
\ibraketop{0\down,1\up}{\Wh}{0\down,1\up}]$
and the higher energy singlet state
$\Lambda_{\Hx,\text{ss}}=\frac{1}{\sqrt{2}}[
\ibraketop{0\up,1\down}{\Wh}{0\up,1\down} +
\ibraketop{0\down,1\up}{\Wh}{0\down,1\up}]$.
Both inherit the correct spin qualities via the diagonalization of
$\mat{W}$.

Finally, we can expand these out to find
\begin{align*}
  \Lambda_{\Hx,\gs}=&
  \int \frac{dxdx'}{2}U(x-x')2\phi_0(x)^2\phi_0(x')^2
  \\
  \Lambda_{\Hx,\ts}=&\int \frac{dxdx'}{2}U(x-x')
         [\phi_0(x)\phi_1(x')-\phi_1(x)\phi_0(x')]^2
  \\
  \Lambda_{\Hx,\ss}=&\int \frac{dxdx'}{2}U(x-x')
         [\phi_0(x)\phi_1(x')+\phi_1(x)\phi_0(x')]^2
\end{align*}
in our specific case, as in
Eqs.~\eqref{eqn:Lambdags}, \eqref{eqn:Lambdats}
and \eqref{eqn:Lambdass}.
The Hx energy is then given by
\begin{align}
  \EHx=w_{\gs}\Lambda_{\Hx,\gs}
  + w_{\ts}\Lambda_{\Hx,\ts} + w_{\ss}\Lambda_{\Hx,\ss}.
\end{align}

\section{The difference between exact and KS densities}
\label{app:Densities}

Equation~\eqref{eqn:nEquiv}, restated here for convenience,
\begin{align}
  \np=&(1-p)n_{\gs} + p n_{\ts}
  =(1-p)n^{(p)}_{s,\gs} + p n^{(p)}_{t,\ts}
  \nonumber\\
  =& (2-p)|\phi_0^{(p)}|^2+p|\phi_1^{(p)}|^2,
\end{align}
shows the relationship between the exact and Kohn-Sham densities, and
the two orbitals that go into the latter. It may be tempting, at first
glance, to assume that $n_{\gs}=n^{(p)}_{s,\gs}$ and
$n_{\ts}=n^{(p)}_{s,\ts}$. As illustrated below this is not
the case in general, and any similarity $n_{\gs}\approx n^{(p)}_{s,\gs}$
and $n_{\ts} \approx n^{(p)}_{s,\ts}$ between the real and KS 
densities highlights a success of the EDFT formalism in retaining an
intuitive understanding of the densities involved.

The latter point is most obvious when we consider a singlet state as
well. We note that the triplet- and singlet- densities of interacting
states are not the same, i.e. $n_{ts}\neq n_{\ss}$ in general
(see, e.g. Figure~\ref{fig:DensS}).
However, as noted in the previous section the corresponding
KS densities are independent of the choice of spins, and
$n_{s,\ts}=n_{s,ss}=|\phi_0|^2+|\phi_1|^2$ are identical.
Ergo, the KS densities cannot be the same as the
interacting densities. In the singlet/triplet case, having
$n_{s,\gs}=|\phi_0|^2=n_{\gs}$ would require, at a minimum, that
$n_{\ts}-n_{\gs}/2=n_{s,\ts}-n_{s,\gs}=|\phi_1|^2>0$, a situation that
cannot be guaranteed in general.

Another perspective to this issue is provided by considering the degrees of freedom available to
the problem. Both $\phi_0$ and $\phi_1$ must, by virtue of the GOK
generalization of the Hohenberg-Kohn theorem, be eigenfunctions of the
same one-body Hamiltonian $\hat{h}_s=\th+\vh_s$, where the multiplicative
potential $v_s$ acts a continuous Lagrange multiplier
that constrains the non-interacting density $n_s$ to be equal to
$n$. Thus $n_{s,\gs}=2|\phi_0|^2$ and
$n_{s,\ts}=|\phi_0|^2+|\phi_1|^2$ come from a constrained problem with
just one continuous Lagrange multiplier, $v_s$, for one continuous
constraint, $(2-p)|\phi_0^2|+p|\phi_1^2|=\np$.
Matching the components of the density $n_{\gs}$ and $n_{\ts}$
separately would require \emph{two} continuous constraints.
But in this case we have \emph{three densities},
$n_{\gs}$, $n_{\ts}$ and $n_{\ss}$, that must be reproduced by just
\emph{two orbitals} coming from a \emph{single potential} $v_s$
-- clearly an impossible task in general.
Quite generally, any new density would require its own Lagrange multiplier.
Hence, given the over-constrained nature of the problem, it is
fortunate and not at all obvious that the KS densities $n_{s,\FE}$ of
components even qualitatively resemble their interacting counterparts
$n_{\FE}$.

%

\end{document}